\documentclass[unnumsec,webpdf,contemporary,large,numbered]{oup-authoring-template}
\usepackage{booktabs,tabularx}
\usepackage{xcolor}
\usepackage{etoolbox}
\usepackage{placeins}
\makeatletter
\def\ps@preprint{%
  \def\@oddhead{\small\sffamily MesoCraft\hfill Preprint}%
  \let\@evenhead\@oddhead
  \def\@oddfoot{\hfil\small\thepage\hfil}%
  \let\@evenfoot\@oddfoot}
\let\ps@opening\ps@preprint
\let\ps@headings\ps@preprint
\patchcmd{\@maketitle}{\vspace*{40bp}}{\vspace*{6bp}}{}%
  {\PackageWarning{mesocraft}{Preprint title spacing patch was not applied}}
\makeatother
\graphicspath{{figures/}}
\newcommand{\societylogo}{}

\begin{document}
\setcitestyle{numbers,round}
\raggedbottom
\pagestyle{preprint}
\journaltitle{}
\DOI{}
\copyrightyear{}
\pubyear{}
\vol{}
\issue{}
\access{}
\appnotes{}

\firstpage{1}
\title[MesoCraft Web]{MesoCraft: a web-based tool for interactive procedural modeling of mesoscale biological structures}
\author[1,$\ast$]{Donggang Jia\ORCID{0000-0002-1358-8718}}
\author[1]{Alexandre Kouyoumdjian\ORCID{0009-0005-0915-369X}}
\author[2,3]{Ondrej Strnad\ORCID{0000-0002-8077-4692}}
\author[4]{Uros Smajdek\ORCID{0000-0002-8127-7700}}
\author[4]{Ciril Bohak\ORCID{0000-0002-9015-2897}}
\author[1]{Dilara Bayraktar\ORCID{0009-0001-1506-9989}}
\author[1]{Alejandra Esparza\ORCID{0009-0001-2356-4509}}
\author[5]{Peter Mindek\ORCID{0000-0002-9434-5952}}
\author[5]{Tobias Klein\ORCID{0000-0001-9455-7587}}
\author[6]{Stefan T. Arold\ORCID{0000-0001-5278-0668}}
\author[1]{Ivan Viola\ORCID{0000-0003-4248-6574}}
\address[1]{\orgdiv{Computer, Electrical and Mathematical Sciences and Engineering Division}, \orgname{KAUST}, \orgaddress{\street{Thuwal}, \postcode{23955}, \country{Saudi Arabia}}}
\address[2]{\orgname{Mohamed bin Zayed University of Artificial Intelligence}, \orgaddress{\street{Abu Dhabi}, \country{United Arab Emirates}}}
\address[3]{\orgname{GenBio AI}, \orgaddress{\street{Palo Alto, CA}, \postcode{94301}, \country{USA}}}

\address[4]{\orgdiv{Faculty of Computer and Information Science}, \orgname{University of Ljubljana}, \orgaddress{\street{Ljubljana}, \postcode{1000}, \country{Slovenia}}}
\address[5]{\orgname{Nanographics GmbH}, \orgaddress{\street{Vienna}, \postcode{1030}, \country{Austria}}}
\address[6]{\orgdiv{Biomedical Sciences Division}, \orgname{KAUST}, \orgaddress{\street{Thuwal}, \postcode{23955}, \country{Saudi Arabia}}}
\corresp[$\ast$]{Corresponding author. \href{email:donggang.jia@kaust.edu.sa}{donggang.jia@kaust.edu.sa}}
\abstract{Biological models bring the molecular world of cells into view, enabling experts and the public to explore and understand biological discoveries. Building mesoscale models requires organizing many molecular components into spatial arrangements that can be revised as evidence and interpretations develop. We present MesoCraft, a browser-based tool for structural biologists centered on interactive procedural modeling. User-defined rules generate collections of molecular instances, while a geometry-node editor and visual feedback support authoring and revision of their spatial relationships. Beyond model construction, MesoCraft integrates exploration and communication, enabling researchers to inspect molecular organization and create guided animations that highlight key spatial relationships. By connecting modeling, exploration, and explanation in a unified workflow, MesoCraft allows insights gained during exploration and communication to inform iterative model refinement, supporting the development of mesoscale models into visual narratives of cellular organization. A demonstrative deployment of the tool is available at \url{https://mesocraft.kaust.edu.sa/}.}
\otherabstract[Graphical abstract]{%
  \centering
  \includegraphics[width=0.8\linewidth]{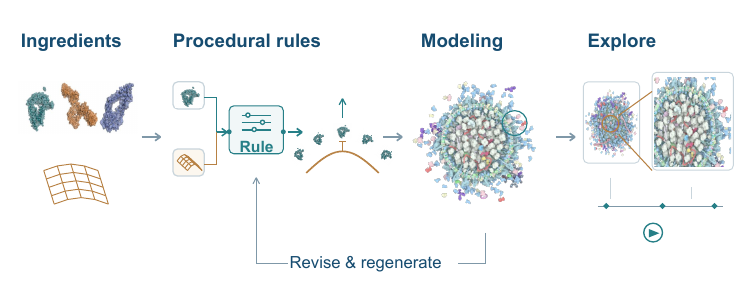}\par
}
\maketitle

\section{Introduction}
Biological discoveries often begin with an image that reveals an unexpected structure or a new view of cellular organization. Connecting these observations to the molecules that make up a biological system requires integrating evidence across scales. Microscopy reveals cellular architecture, whereas structural biology provides detailed molecular structures. Cryo-electron tomography (cryo-ET) brings these perspectives closer by visualizing macromolecular assemblies within cells; combined with subtomogram averaging, it can resolve selected complexes at high resolution, as demonstrated for eukaryotic ribosomes in situ \citep{hoffmann2022insitu}. Yet images and reconstructions do not generally identify and locate every molecular component in a heterogeneous cellular region. Building a molecular account of that region therefore requires combining imaging observations with known structures, composition estimates and spatial information \citep{johnson2015cellpack,nguyen2021mesocraft}.

Mesoscale models turn this combined evidence into spatial representations of molecules in their spatial context, for example as macromolecular assemblies, or within membranes and other cellular compartments. Constructing mesoscale models requires decisions about which components to include and how to arrange them where evidence is incomplete. Dedicated modeling tools make these decisions explicit and editable, helping researchers explore alternative interpretations and communicate them to experts and the public \citep{johnson2015cellpack,nguyen2021mesocraft}.

Several biological modeling tools address this integration. CellPACK~\citep{johnson2015cellpack} uses composition recipes and packing algorithms to construct three-dimensional molecular environments. Mesoscope provides a web-based environment for integrating and curating molecular ingredients, their representations and portable recipes for mesoscale modeling \citep{autin2020mesoscope}. CellPAINT~\citep{gardner2018cellpaint,gardner2021cellpaint} provides a complementary, direct authoring approach in which molecular brushes support the interactive illustration of cellular environments. Klein et al.~\citep{klein2018instant} combined construction and visualization to enable interactive population of crowded biological environments.

At larger scales, authoring must also accommodate the number of molecular copies and the cost of revising their arrangements. Procedural modeling expresses spatial relationships as reusable rules: a component can be placed relative to a reference surface, distributed within a volume or repeated relative to another component. A single rule generates a collection of instances, so revisions can be made to the construction specification rather than to each copy separately. Interactive construction has already been demonstrated in prior work \citep{klein2018instant,nguyen2021mesocraft}; a continuing challenge is to make expressive rule authoring accessible alongside visual inspection and presentation of the resulting organization.

Generating large collections of instances also increases the atomic detail that visualization must handle. Mol* Viewer~\citep{sehnal2021molstar} demonstrates the importance of high-performance graphics and data management for viewing large biomolecular structures in the browser, while MolViewSpec~\citep{midlik2025molviewspec} uses declarative descriptions to make molecular views, annotations and visual narratives reproducible and reusable. These capabilities complement spatial model construction. During procedural authoring, responsive rendering is needed to inspect the effects of modeling decisions; subsequently, selected views and visual sequences help explain the organization to others. Connecting these tasks requires continuity between the editable construction and the scene being explored or presented.

In this work we present MesoCraft, a browser-based environment for authoring mesoscale organization through explicit procedural rules. By retaining the construction alongside its visual result, the system connects model revision to the exploration and communication of molecular arrangements. We describe this workflow through the modeling tools, representative scenes and their use in visual explanations.

\section{System overview}
\begin{figure*}[!t]
\centering\includegraphics[width=\textwidth]{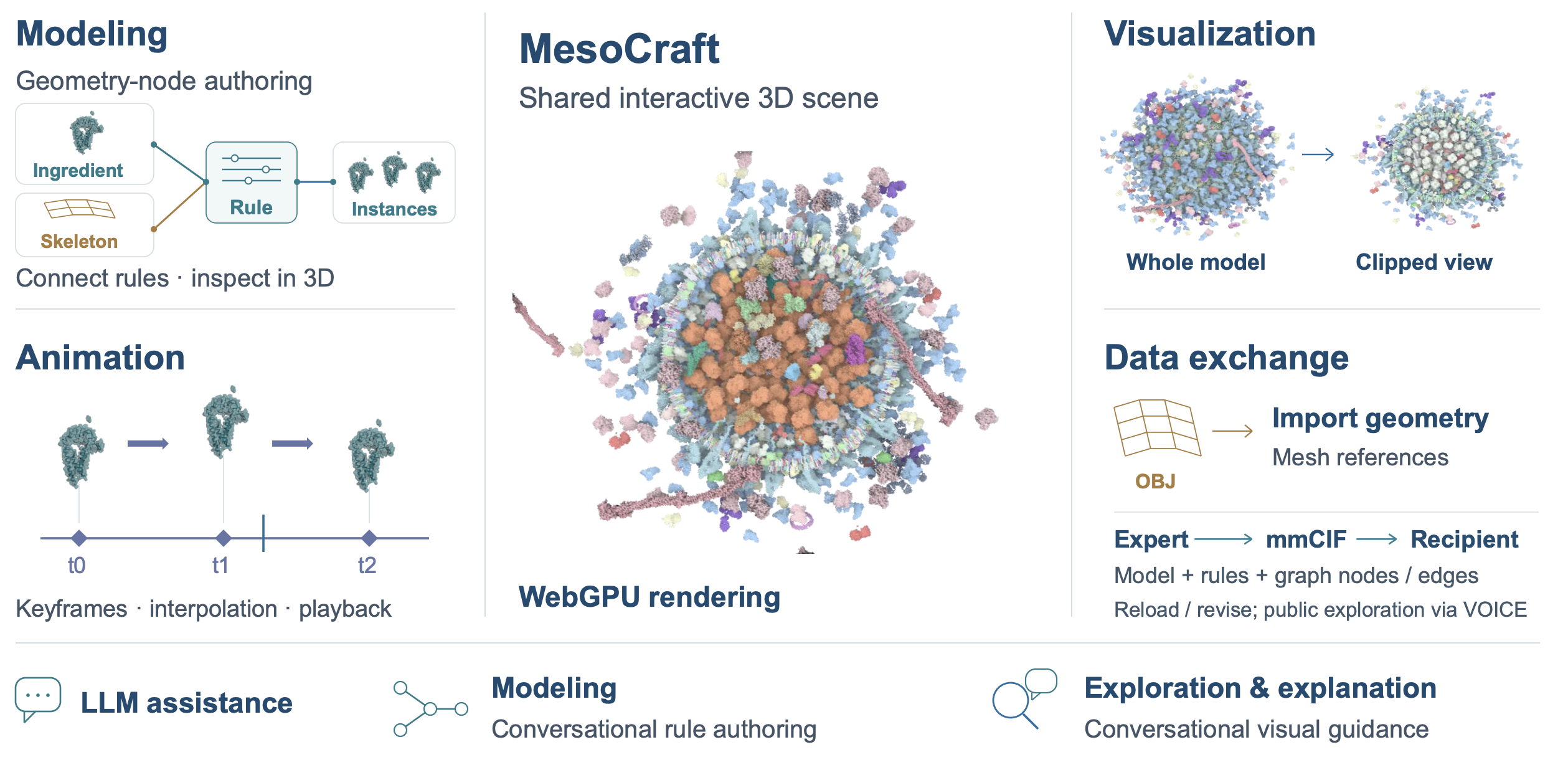}
\caption{MesoCraft system overview. Procedural modeling, visualization and animation share a three-dimensional scene. Geometry-node authoring connects ingredients and references to rules and generated instances. Visualization supports clipping, coloring and labels; animation provides keyframes, interpolation and playback. OBJ supplies geometry references. Experts share enriched mmCIF files containing models, rules and node graphs: recipients can reload and revise the construction, while public audiences can explore the loaded model through VOICE. Conversational assistance also supports rule authoring.}
\label{fig:overview}
\end{figure*}

MesoCraft brings graph-based procedural modeling, visualization and animation into a shared browser workspace (Figure~\ref{fig:overview}). Modeling forms the core: users connect molecular ingredients and geometric references to rules that generate spatial arrangements. The resulting scene supports inspection during construction and subsequent presentation. The graph retains the relationships used to build the model, allowing researchers to return to them as their interpretation develops.

\begin{figure*}[!t]
\centering
\includegraphics[width=\textwidth]{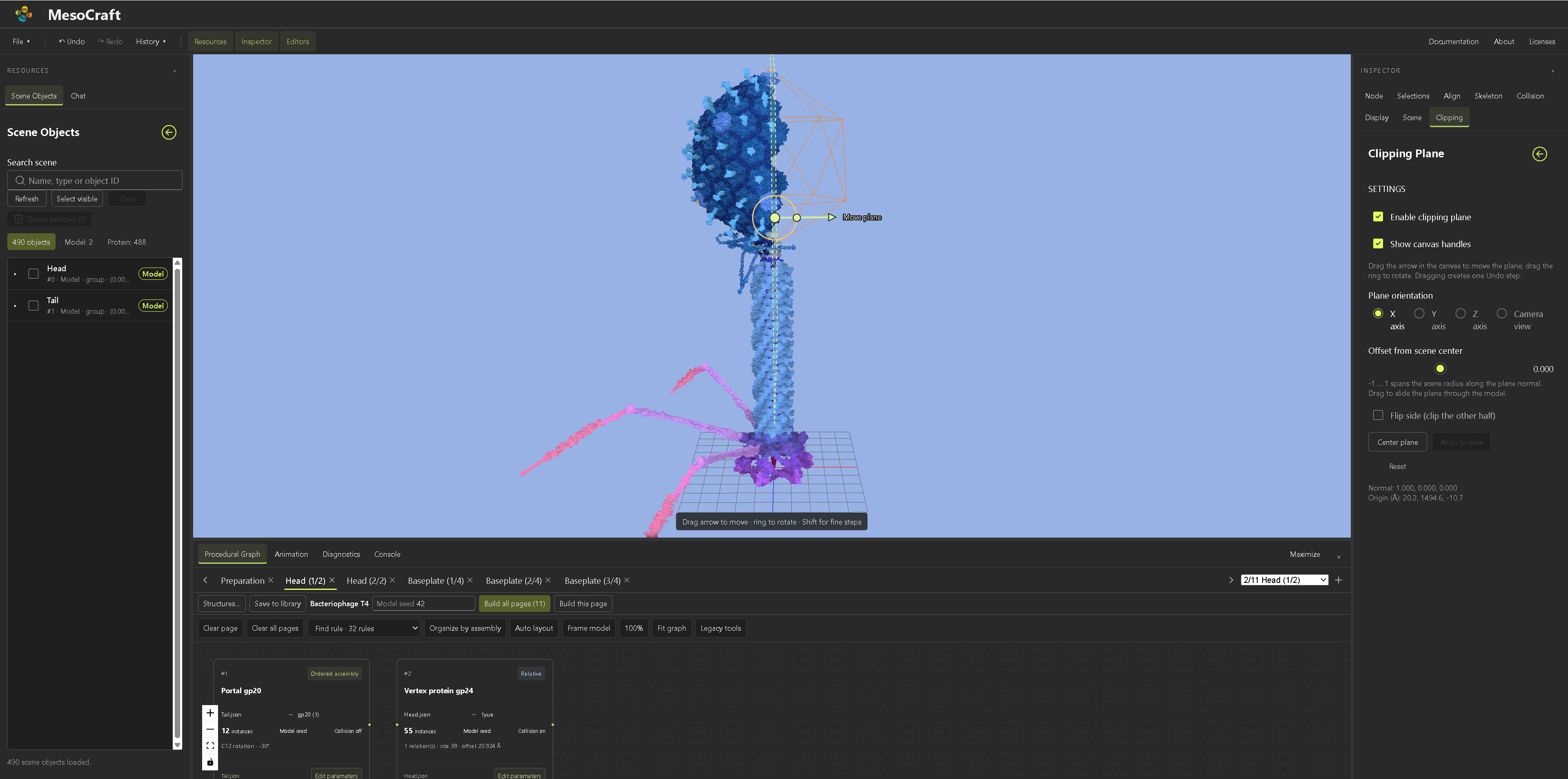}
\caption{MesoCraft Web interface after loading and building the T4 bacteriophage graph. The left panel shows the scene hierarchy, with the head and tail assemblies and their object counts. The central viewport displays the assembled T4 bacteriophage model and an active clipping gizmo. The right inspector exposes clipping-plane orientation, offset and visibility controls, while the lower panel shows procedural-graph pages and selected rule cards, including generated instance counts. The screenshot illustrates how hierarchy, rule authoring, clipping and visual inspection remain visible in a single browser workspace.}
\label{fig:t4-interface}
\end{figure*}

The workspace makes the construction and its results accessible together (Figure~\ref{fig:t4-interface}). The hierarchy organizes assemblies and their components, the graph panel exposes construction relationships, and the central viewport displays the scene. The Inspector groups controls for display and color, alignment, skeletons, collisions and clipping. Navigation and labels support exploration across scales. These complementary views connect a selected structure to its place in the assembly and to the rules responsible for its arrangement.

Animation extends this exploration into sequences that direct attention and present changes over time. Its timeline editor organizes actions on selected components using the same model that users inspect. VOICE~\cite{jia2025voice} provides conversational exploration and visual explanations, while Chat Modeling supports natural-language authoring. Enriched mmCIF sharing carries the scene and its editable construction to other users, who can continue modeling or explore it through their own questions. The following sections follow this progression from constructing an arrangement to examining, presenting and sharing it.

\section{Procedural modeling}
\subsection{Ingredients, skeletons and generated instances}
MesoCraft expresses a model as reusable molecular \textbf{ingredients}, geometric references called \textbf{skeletons}, construction \textbf{rules} and the \textbf{instances} produced by those rules. An ingredient supplies molecular coordinates; an instance supplies a placement of a copy. A skeleton serves as a geometric reference for defining the positions and orientations of molecular instances. A rule combines these inputs with user-defined spatial relations.

We write a placement as $(\mathbf{x},R)$, mapping an ingredient-local point $\mathbf{v}$ to $\mathbf{x}+R\mathbf{v}$ in world coordinates. If a rule places a selected ingredient pivot $\mathbf{p}$ at a target $\mathbf{q}$, its translation must account for the rotation:
\begin{equation}
 \mathbf{x}=\mathbf{q}-R\mathbf{p}.
 \label{eq:pivot}
\end{equation}
This distinction explains why placing a molecule's center and placing a selected attachment point are different operations. The rule descriptions below explain how users combine these placement settings with geometric references to construct different arrangements.

\subsection{Skeletons: describing biological shapes and spatial regions}
\begin{figure*}[!t]
\centering\includegraphics[width=\textwidth]{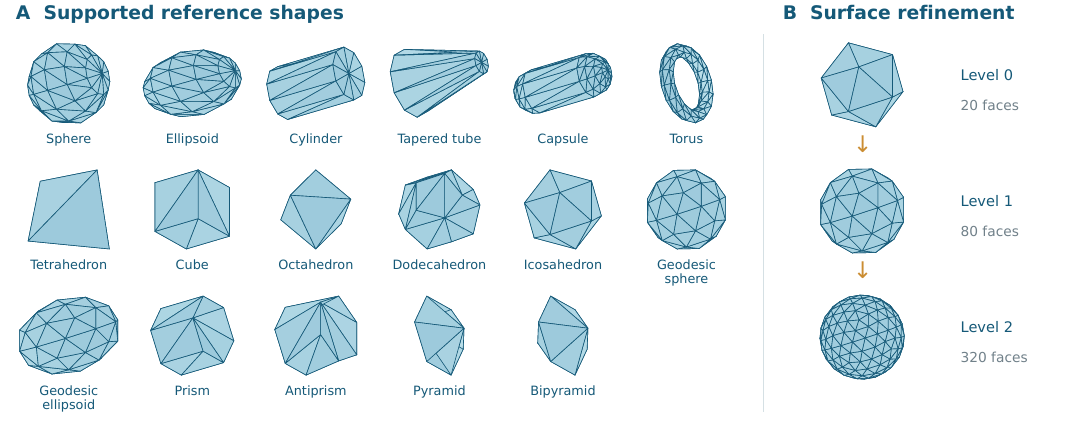}
\caption{Skeletons as geometric references for procedural modeling. (A) The 17 shape types available in the built-in generator, shown at independent scales; axial polygonal examples use six sides. (B) Different levels of surface refinement provide control over geometric detail. Skeletons describe the shape or spatial extent of a modeled structure, while rules specify how molecular instances occupy or relate to that geometry.}
\label{fig:skeleton-generator}
\end{figure*}
Skeletons describe the overall shapes of biological structures and delineate regions within a mesoscale model. A surface can represent a membrane boundary, while an enclosed volume can define a compartment or a region to be populated with molecular components. Multiple skeletons allow researchers to organize a scene into regions with different compositions or placement rules. These geometric references connect a model's overall spatial organization to the arrangement of its molecular ingredients.

The built-in generator lets users create geometric references directly in the browser (Figure~\ref{fig:skeleton-generator}). Spheres and ellipsoids can describe the envelopes of vesicles or rounded compartments, while cylindrical and tapered shapes can describe the spatial extent of fibers or elongated assemblies. Users adjust radii and axial dimensions to match the size and proportions of the structure being modeled, establishing boundaries for surface placement or regional filling. Surface refinement controls the level of geometric detail. For more complex biological shapes, users can import OBJ meshes prepared in other tools.

Once the reference geometry is prepared, users connect it to rules that determine molecular placement and orientation. A single skeleton can support several component populations, providing a common spatial reference as the construction develops.

\subsection{Statistical distributions, orientation and placement constraints}
Spatial relationships can be specified using fixed values or statistical distributions, allowing users to combine regular arrangements with controlled variability. For distance settings, users can choose a preferred value and a spread, with uniform or Gaussian sampling where supported. A fixed distance produces a consistent offset, while a distribution allows molecular copies to occupy a range of distances, for example around a reference surface. Angular variation provides similar control over molecular orientation. These settings let researchers express spatial variability rather than assigning the same position or rotation relationship to every copy.

Orientation controls determine how ingredients relate to their surroundings. Users can specify a rotation, randomize orientations or align ingredients to skeleton geometry. For example, alignment to a local surface normal can orient membrane-associated components relative to a curved boundary, with an option to reverse their direction. Tangent alignment and rotation about the aligned axis, where supported, provide further control over the arrangement within the surface. Angular perturbations can introduce variation around a preferred orientation.

Placement constraints help users control which arrangements are admitted to the model. Depending on the rule, users can enable collision checks, require a minimum distance between molecular centers or restrict placement to a specified region. These controls support the construction of crowded scenes while limiting overlaps and keeping components within their intended compartments. Tight constraints can reduce the number of instances that can be placed, so users inspect the resulting organization and adjust spacing, distributions or regional boundaries as needed.

\begin{figure*}[!t]
\centering\includegraphics[width=\textwidth]{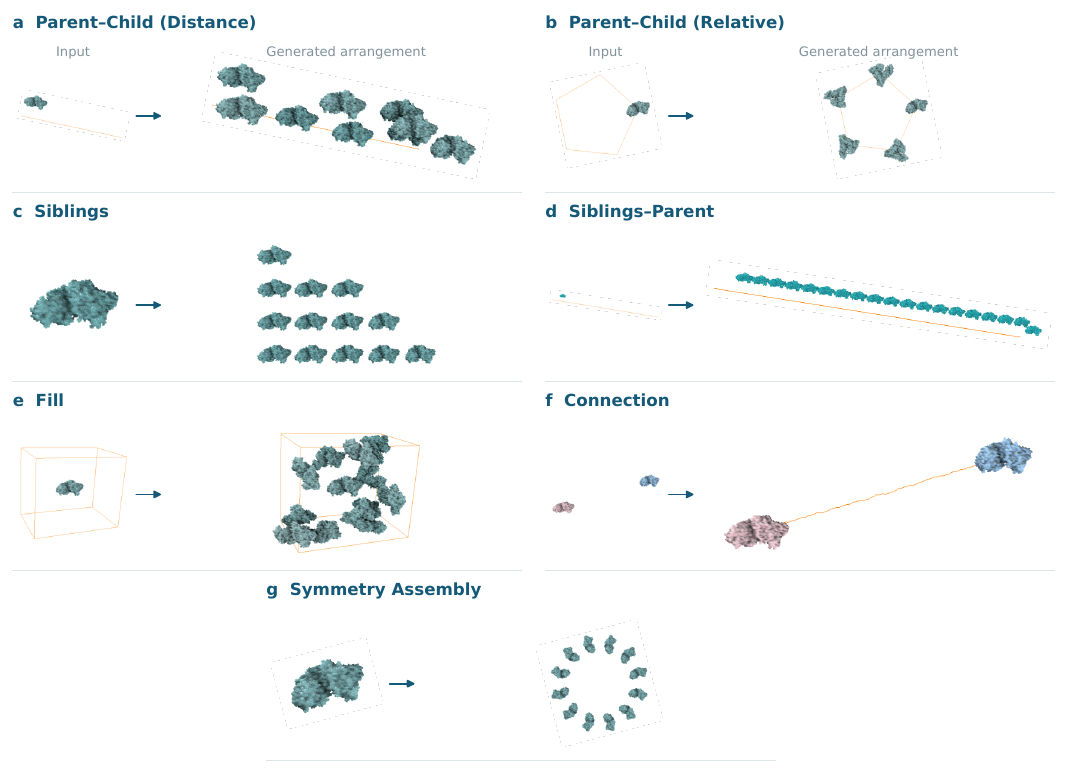}
\caption{Rule types illustrated by before-and-after views: (a) Parent--Child Distance, placing copies relative to a reference surface; (b) Parent--Child Relative, repeating a relative placement at corresponding scaffold locations; (c) Siblings, extending relationships between neighboring instances; (d) Siblings--Parent, combining neighbor relationships with a skeleton reference; (e) Fill, populating a region; (f) Connection, generating a path between selected copies of the same ingredient; and (g) Symmetry Assembly, generating a ring of 12 instances. Inputs and generated arrangements are independently scaled for visibility; within each pair, the view orientation is consistent.}
\label{fig:rules}
\end{figure*}

\subsection{Rule types and their modeling applications}
Building on the statistical and rule-based modeling framework of Nguyen et al.~\cite{nguyen2021mesocraft}, MesoCraft extends the original rule set with additional modeling operations and geometric controls. The seven rule types address complementary spatial relationships (Figure~\ref{fig:rules}). Users choose a rule according to whether a component should follow a reference shape, repeat relative to neighboring components, occupy a region or form a symmetric assembly. The examples in Figure~\ref{fig:rules} use a shared molecular ingredient to illustrate these relationships independently of a particular biological system.

\textbf{Parent--Child (Distance):} This rule places molecular copies at a specified distance from a skeleton (Figure~\ref{fig:rules}a). It is useful for distributing membrane-associated proteins around a vesicle or positioning components in a layer near a reference surface. Users select the ingredient and skeleton, set the target count and distance or distance distribution, and control which side of the surface the molecules occupy and how they are oriented.

\textbf{Parent--Child (Relative):} This rule repeats a defined relative placement at corresponding locations on a skeleton (Figure~\ref{fig:rules}b). It supports arrangements in which components occupy recurring geometric locations, such as equivalent positions on a polyhedral scaffold. Users define the component's relative position, offset and rotation with respect to the reference geometry. The placement relationship can then be reused across the scaffold, helping maintain a consistent local arrangement as the reference shape changes.

\textbf{Siblings:} This rule extends an arrangement from an existing molecular instance using relationships between neighboring copies (Figure~\ref{fig:rules}c). It is suitable for constructing chains, filaments or repeated molecular arrays whose organization is described by local interactions. Users control the relative displacement and rotation between neighbors, the number of copies and any supported statistical variation. Multiple neighbor relationships allow an arrangement to extend in more than one direction.

\textbf{Siblings--Parent:} This rule combines relationships between neighboring molecules with a shared skeleton reference (Figure~\ref{fig:rules}d). It is useful when a repeated arrangement must follow an underlying shape, for example when components extend along a curved scaffold while maintaining an offset from it. Users specify the neighbor relationship, the reference skeleton and the desired distance and orientation relative to that reference. This couples local molecular organization to the larger biological shape.

\textbf{Fill:} This rule populates an enclosed region or distributes components along a path (Figure~\ref{fig:rules}e). Volume filling can represent molecular crowding within a compartment, while path filling can construct a sequence of molecular copies along a filament reference. Users choose the ingredient and target geometry, set the requested count or supported spacing controls, and configure orientation and placement constraints. Regional boundaries and collision settings help control the resulting occupancy.

\textbf{Connection:} This rule creates a geometric path between selected components (Figure~\ref{fig:rules}f). It supports the construction of references for linkers or connecting fibers between assemblies. Users choose the endpoints and attachment positions, the connection mode and, where supported, a target path length. The resulting path can serve as a skeleton for subsequent placement rules, allowing a connection to be populated with molecular ingredients.

\textbf{Symmetry Assembly:} This rule generates regular arrangements from initial placements using specified rotations and translations (Figure~\ref{fig:rules}g). It is suitable for protein rings, repeated axial assemblies and helical structures; the figure illustrates a ring of 12 copies. Users control the symmetry axis and center, rotation angle, translation step and repetition count, or select available scaffold sites. These parameters make the intended global organization explicit, supporting the construction and revision of assemblies with repeated geometric relationships.

\subsection{Interactive construction and revision}
The graph panel lets users organize a model as a sequence of connected construction steps (Figure~\ref{fig:t4-interface}). Connections show which ingredients and geometric references feed a rule and where its generated instances enter the construction. Users can follow these relationships to understand how an arrangement was produced and locate the rule they want to revise. A shared skeleton can guide several rules, while generated instances or connection paths can provide references for subsequent steps. This makes the graph a working description of how individual placements combine into a larger model.

For example, a user constructing a compartment can begin with an ellipsoidal skeleton to define its boundary and load ingredients for surface-associated and internal components. A Parent--Child Distance rule distributes the surface-associated ingredient relative to that boundary, with orientation controls aligning it to the local surface. A Fill rule then populates the enclosed region with internal components, using the same geometry to relate the interior to the surrounding boundary. After inspecting the combined scene, the user can adjust the surface offset or orientation, revise the filling count or spacing, and apply the edited rules to inspect the updated arrangement. Collision and regional constraints help organize the components as these steps are combined. The example illustrates how a few complementary rules connect overall shape, surface organization and interior composition.

As a construction grows, users can group related steps into graph pages for different assemblies or modeling stages. This keeps the relevant ingredients, references and rules together while the scene hierarchy provides an overview of the resulting objects. Users can inspect an individual rule, change its parameters, apply it and revert its result during revision. Working through related steps in the graph helps them identify which later placements use a changed reference and review those steps in turn. The three-dimensional view provides immediate context for judging each revision, while the graph preserves the relationships needed to continue developing the model.

\section{Models constructed with procedural rules}
\begin{figure*}[!t]
\centering\includegraphics[width=\textwidth]{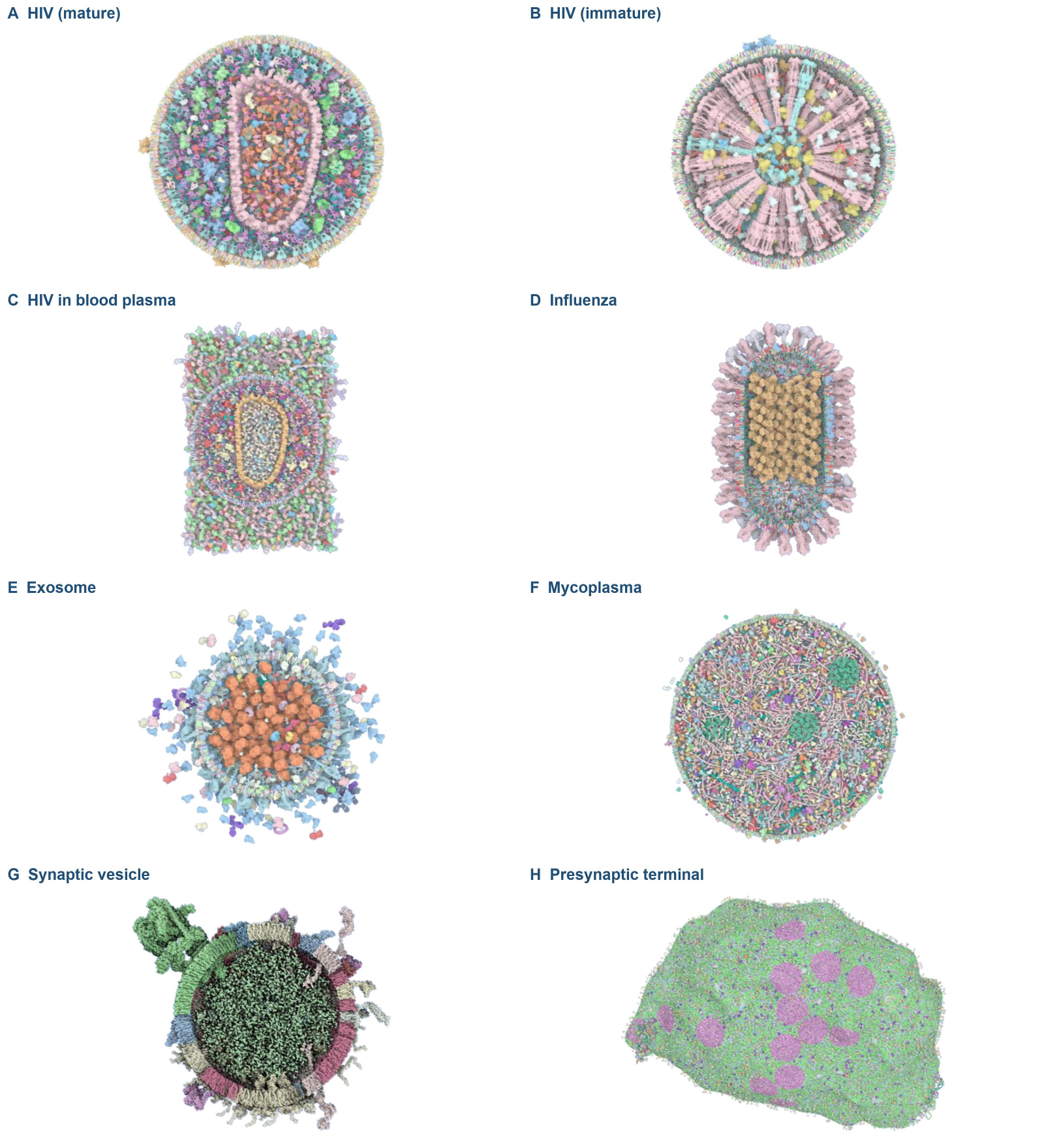}
\caption{Mesoscale scenes constructed with procedural rules, with non-procedural models from the Mesoscope example collection as references~\citep{autin2020mesoscope}. (A) Mature HIV, 141 rules. (B) Immature HIV, 89 rules. (C) Mature HIV in blood plasma, 152 rules. (D) Influenza, 45 rules. (E) Exosome, 117 rules. (F) Mycoplasma, 1,344 rules. (G) Synaptic vesicle, 31 rules. (H) Presynaptic terminal, 84 rules. Counts indicate the number of MesoCraft rule entries used in each construction. The Mycoplasma reference is the curated model of Maritan et al.~\citep{maritan2022mycoplasma}. SARS-CoV-2 and T4 bacteriophage are discussed separately in the case studies. Panels are independently framed without a common scale.}
\label{fig:examples}
\end{figure*}

Figure~\ref{fig:examples} shows how combinations of rules express organization across particles, vesicles, a cellular environment and a presynaptic terminal. These scenes were constructed in MesoCraft using non-procedural models from the Mesoscope example collection as references~\citep{autin2020mesoscope}. The gallery brings together surface organization, populated interiors and repeated assemblies within complete models; individual sources and rule counts appear in the caption.

The examples contain 31--1,344 rule entries. A rule entry specifies a construction operation that may generate many molecular instances. The cutaway views reveal the internal organization that users can examine with the visualization tools introduced next.

\section{Visual exploration and explanation}
Visual exploration helps users assess the organization produced by their rules. Moving between the complete assembly and a selected region reveals how local arrangements contribute to the larger structure. Component colors provide visual references across these views, while screen-space ambient occlusion conveys depth in crowded regions.

An inspection can begin with a spatial question: how does a selected population occupy its compartment, and how does it relate to the surrounding boundary? The user locates that population in the hierarchy, examines it at closer range and moves the clipping plane through the scene to expose successive regions. Labels help distinguish nearby component types, while the remaining boundary provides context for interpreting their placement. These observations help users identify which construction relationships to revisit in the graph. Views that clarify the relationship also help the author plan a presentation: an overview establishes the setting, and a selected interior view identifies the region that an animation or conversational explanation can subsequently address.

\subsection{Labeling crowded molecular scenes}
MesoCraft's labeling builds on Labels on Levels~\citep{kouril2019labels} and HyperLabels~\citep{kouril2021hyperlabels}. These methods use scene hierarchy and representative annotations to support exploration of crowded, repeated molecular structures. The WebGPU implementation selects representative type-level labels using visible regions, depth and screen-space information, connecting component names to the displayed model.

Label counts and spacing adapt to viewing distance, while screen-space adjustment and conflict removal reduce overlap. Users can adjust font size, label budgets and spacing for an overview or detailed inspection. Labels identify representative components according to visibility and available layout space, complementing the colors assigned to component classes. Camera and clipping changes update which representatives are shown.

\section{Animation as an editable visual explanation}
\subsection{Selecting objects and organizing a timeline}
To develop an inspected scene into an animated explanation, users first choose the components to animate through a timeline's target filter, which can identify a molecular type, selected instances or another supported subset of the scene. Other components provide spatial context. Separate timelines allow different groups to receive distinct actions and timing, so an explanation can focus on one part of an assembly before moving to another.

Users add keyframes at chosen times, select an animation action and adjust its parameters. Keyframe timing determines when a change occurs, while action settings determine its direction, extent or appearance. Transitions can be configured between compatible states, and time-driven actions provide continuous or repeating motion. Scrubbing and playback let users examine intermediate views, adjust the pace and check that the intended components remain understandable throughout the sequence.

\subsection{Animation actions and presentation presets}
The action library supports translation, rotation about a selected axis or center, continuous spin and orbital motion. These actions can reposition or reorient components to reveal spatial relationships. Radial expansion separates components around a common center, and an explode/reassemble preset provides a sequence that opens an assembly and returns it to its original arrangement. Users control quantities such as displacement, rotation axis, angle, expansion and duration according to the selected action.

Oscillation and rocking provide repeated translational or angular motion, with controls for amplitude and period. Staggered actions introduce delays between instances to show a sequence of changes. Curve-based actions and a multi-curve transport preset let selected instances follow specified paths. Additional actions support morphing and geometric layouts, including planes, rings, spheres and tubes, as well as Brownian-style translation and rotation. These choices provide visual ways to illustrate organization and motion without requiring users to edit each intermediate frame.

Color actions draw attention to selected components while retaining the rest of the scene as context. Users can combine appearance changes with movement on separate timelines, for example highlighting a component before separating it from an assembly. Presets for continuous spin, orbit and eased translation provide starting points that can be adjusted for a particular model and explanation.

\section{Sharing editable models and conversational exploration}
\subsection{Sharing a model for continued editing}
After constructing and inspecting a model, an expert can export it as an enriched mmCIF file for others to open in MesoCraft (Figure~\ref{fig:overview}). The shared file stores the assembled scene together with the rules and node graph saved for that model. Recipients can therefore examine both the molecular organization and the construction relationships that produced it, providing a starting point for discussion and further modeling.

Other researchers can adjust ingredients, geometric references or rule parameters, inspect the revised arrangement and save their own version. The shared graph provides a starting point for exploring alternative organizations and discussing their spatial assumptions. Editable animations can accompany the scene so that collaborators can also adapt its presentation. Supplementary Sections S1--S4 describe authoring and exchange.

\subsection{Public exploration through VOICE}
The same model can be shared with audiences who do not author procedural graphs. After loading an expert-prepared model, members of the public can use VOICE~\citep{jia2025voice} to ask questions about its components and organization and request visual explanations that bring relevant structures into view. Their questions guide the exploration, allowing them to follow their interests and examine parts of the model beyond the views selected for a figure or video. This connects expert model preparation with interactive communication of biological structures.

For users who want to proceed from exploration to modeling, Chat Modeling supports authoring requests expressed in natural language~\citep{jia2026chat}. Users can inspect the resulting actions in the graph and scene interfaces and continue refining the model through direct interaction. Supplementary Sections S5--S6 describe assistant integration.

\section{Case demonstrations}
The following cases follow two paths from procedural construction to communication: an animated presentation of the T4 bacteriophage model (Figure~\ref{fig:animation}) and public exploration of a shared SARS-CoV-2 model through VOICE (Figure~\ref{fig:caseviews}).

\subsection{T4 bacteriophage: graph reconstruction and infection-initiation cue}
\begin{figure*}[!t]
\centering\includegraphics[width=\textwidth]{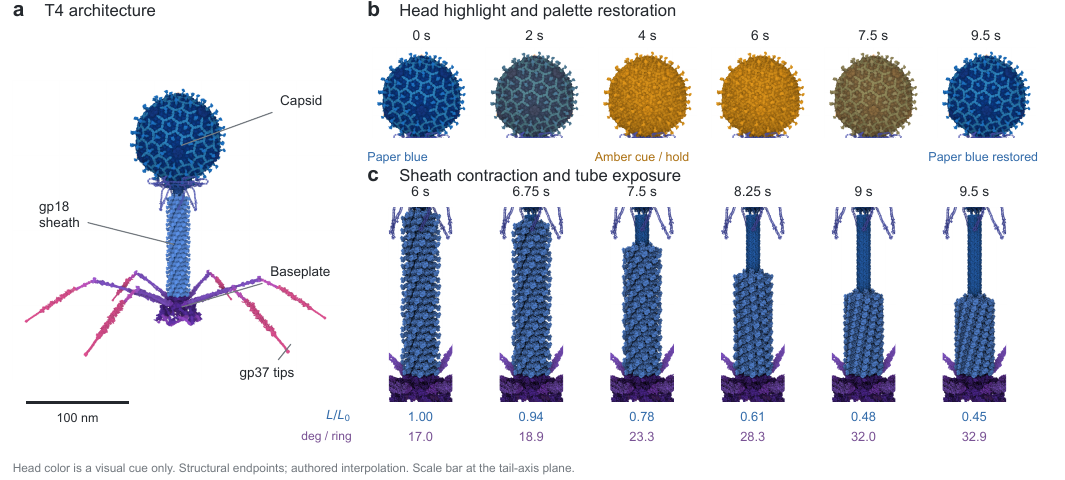}
\caption{T4 bacteriophage graph-to-animation explanation. (a) The complete 488-protein scene in its paper palette, with diffuse matte illumination; the scale bar is calibrated at the tail-axis plane. (b) Fixed-crop head views show the paper blue changing to amber at 1--4~s, holding until 6~s and returning during contraction. The temporary amber highlight directs attention to the head. (c) Six fixed-view tail close-ups show sheath shortening, retwisting and tube exposure. Rows give the authored axial scale $L/L_0$ (ring-center span) and twist per ring in degrees. Structural reports guide the endpoints~\citep{leiman2010t4tail}; timing and transitions are set for visual explanation.}
\label{fig:animation}
\end{figure*}
The T4 bacteriophage case uses the model in Figures~\ref{fig:t4-interface} and~\ref{fig:animation}a~\citep{nanovisT4}. Its editable graph contains 32 rule entries and two guiding skeletons. The graph rebuilds a 488-protein scene: an icosahedral head, portal and neck rings, the tail tube and gp18 sheath helices, a multicomponent baseplate and six long tail fibres. The graph keeps the rotational, helical, path-following and native-assembly relationships accessible for inspection and revision.

The model organization also gives the author a starting point for planning the explanation. The head and tail groups help locate the relevant components, and the author can target the sheath separately from the inner tube when preparing the animation. Keeping these targets distinct allows the sheath to change while the tube remains a spatial reference. Color and movement actions can be assigned to these component groups on separate timelines, coordinating the shift in attention with the structural transition. The model hierarchy thus supplies the objects of the explanation, while the timeline establishes the order in which the audience encounters them.

The animation turns the constructed model into a guided visual explanation (Figure~\ref{fig:animation}). It begins with the complete bacteriophage, establishing the spatial relationship between the head, tail and fibres. A temporary color highlight draws attention to the head before the sequence shifts to the tail, where sheath shortening and twisting reveal the inner tube. Close-up views make this structural change easier to follow, while consistent framing and component colors help viewers relate it to the complete model.

\subsection{SARS-CoV-2: from expert modeling to public exploration}
\begin{figure*}[!t]
\centering
\begin{minipage}[t]{0.32\textwidth}
\centering\textbf{A\quad Initial model}\par\smallskip
\includegraphics[height=42mm]{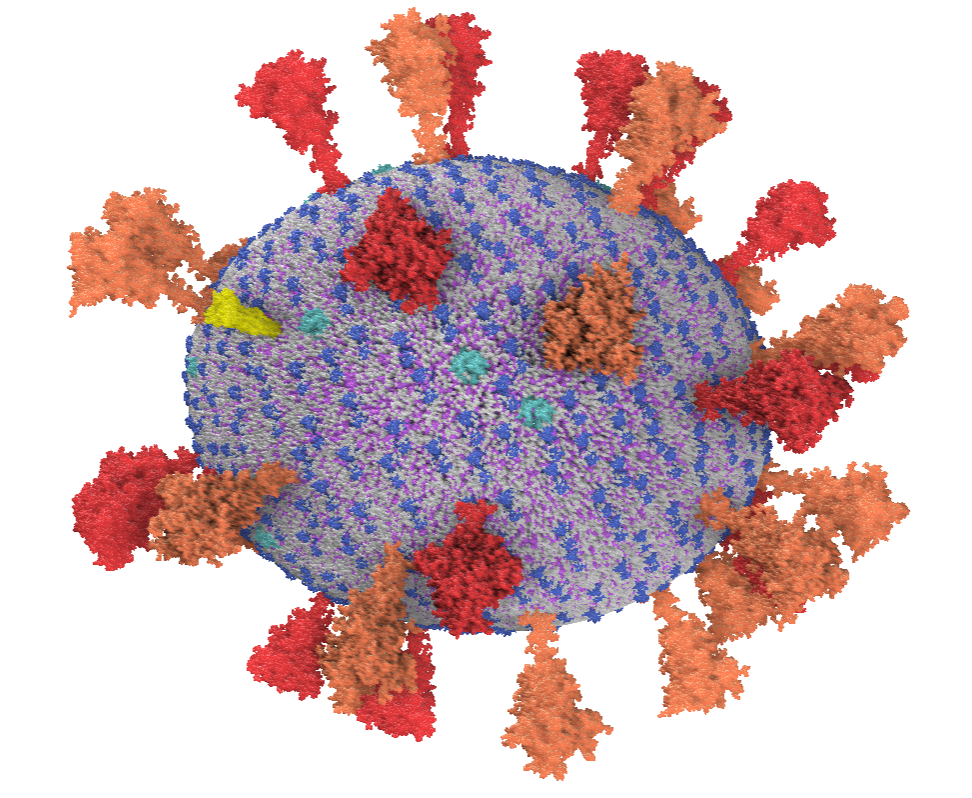}
\end{minipage}\hfill
\begin{minipage}[t]{0.32\textwidth}
\centering\textbf{B\quad Show me RNA}\par\smallskip
\includegraphics[height=42mm]{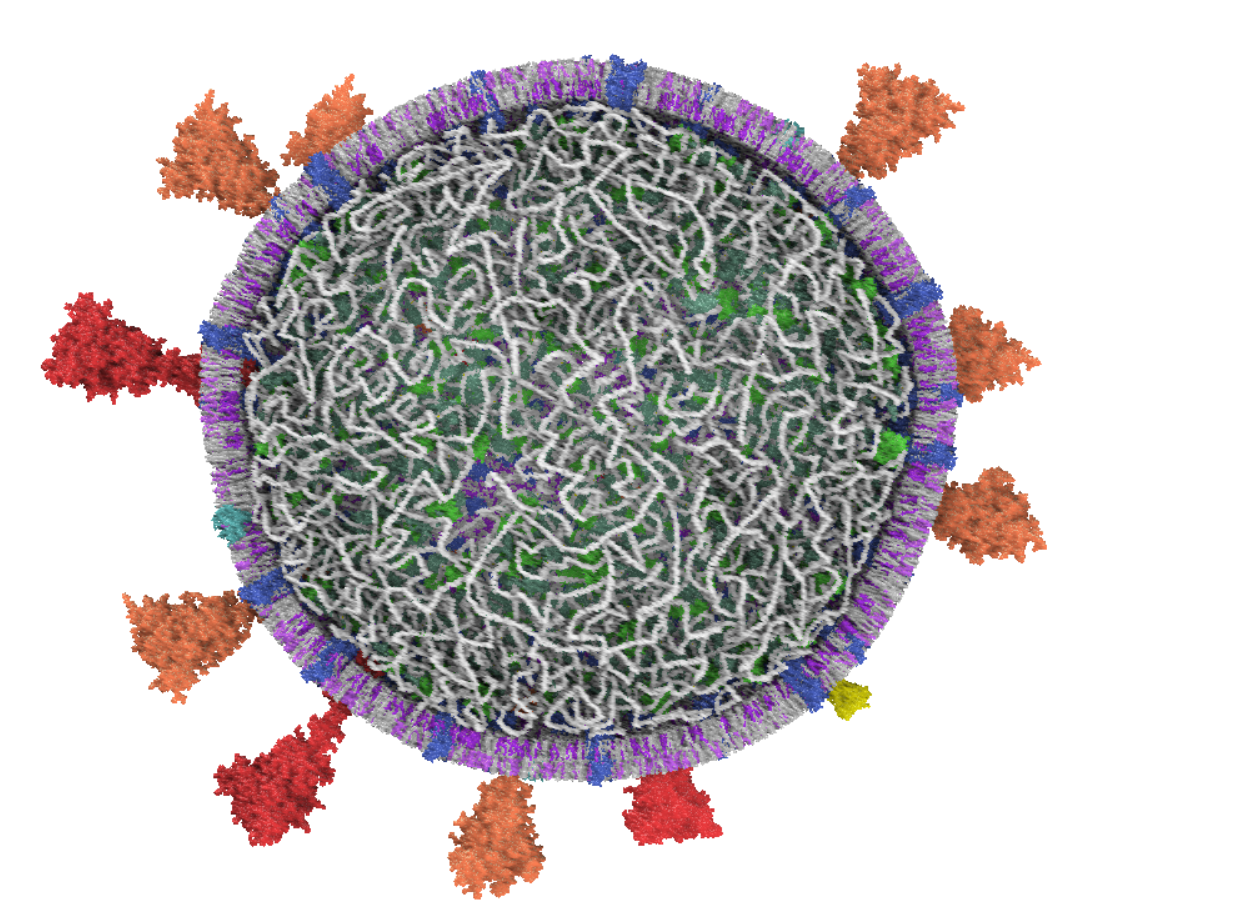}
\end{minipage}\hfill
\begin{minipage}[t]{0.32\textwidth}
\centering\textbf{C\quad What is membrane protein?}\par\smallskip
\includegraphics[height=42mm,trim=415bp 0bp 415bp 0bp,clip]{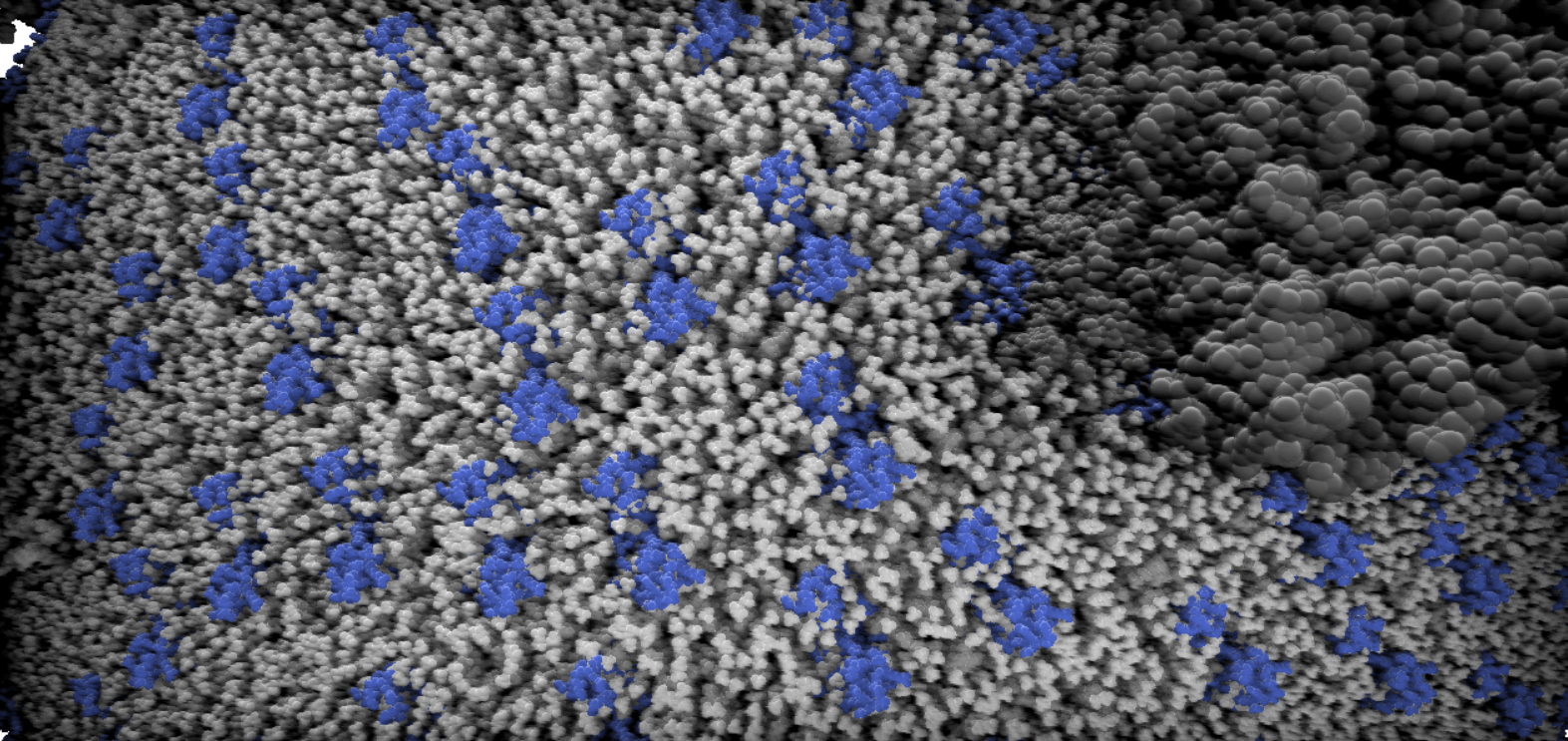}
\end{minipage}
\caption{SARS-CoV-2 demonstration: from an expert-prepared model to conversational exploration through VOICE. (A) The initial model loaded for exploration. (B) Visual output for ``Show me RNA'': a cutaway reveals the white RNA structure within the particle. (C) Visual output accompanying the textual explanation for ``What is membrane protein?'': membrane proteins are highlighted in blue against surrounding structures in gray. Framing and scale are independent.}
\label{fig:caseviews}
\end{figure*}

In the SARS-CoV-2 case, an expert uses procedural modeling to organize molecular components into the scene shown in Figure~\ref{fig:caseviews}A. The shared model provides the starting point for two questions addressed through VOICE~\citep{jia2025voice}.

The request ``Show me RNA'' focuses the exploration on a component inside the particle. Figure~\ref{fig:caseviews}B shows the resulting RNA-focused cutaway. The white RNA structure becomes visible within the surrounding molecular organization, while the envelope and surface components retain the particle context. This view lets a visitor relate an otherwise hidden component to the overall model through a direct conversational request.

A second question, ``What is membrane protein?'', combines a textual explanation with a visualization of the relevant components in the loaded model. In Figure~\ref{fig:caseviews}C, membrane proteins are highlighted in blue while surrounding structures appear in gray. This combination connects the explanation to the component's location in the scene: the user can read the explanation while examining the corresponding structure. Together, the two questions illustrate complementary uses of VOICE---locating a named component and explaining it in its spatial context.

Together, the cases demonstrate two ways to communicate a procedural model: the T4 bacteriophage animation guides viewers through a selected structural transition, whereas the SARS-CoV-2 interactions let visitors choose a component and explore it through questions. Both connect the explanation to identifiable components within the assembled scene.

\section{Web implementation}
The MesoCraft Web frontend is built with React and provides the interface for rule authoring, visualization and animation. Its WebGPU rendering engine is written in C++ and compiled to WebAssembly using Emscripten, enabling it to run in the browser. The React interface sends commands to the WebAssembly application to update the model and its displayed scene. A Node/Express backend provides server-side services, including those used by the conversational assistants. The application requires a browser with support for its WebGPU implementation.

\section{Conclusion}
MesoCraft makes procedural construction the starting point for an ongoing process of model revision, exploration and communication. Keeping spatial relationships editable alongside their visual result enables researchers to develop mesoscale models into explanations that others can examine and extend. This continuity supports collaboration among modelers and gives broader audiences an interactive route into molecular organization.
\section{Acknowledgements}
This work was supported by King Abdullah University of Science and Technology (KAUST) under Grant BAS/1/1680-01-01. The research by S.T.A. reported in this publication was supported by funding from the KAUST Center of Excellence for Smart Health (KCSH), under award number 5932.

\bibliographystyle{mesocraft-nar}
\FloatBarrier
\bibliography{mesocraft-references}
\end{document}